# Towards Kinetic Modeling of Global Metabolic Networks with Incomplete Experimental Input on Kinetic Parameters

Ping Ao[1,2], Lik Wee Lee[1], Mary E. Lidstrom[3,4], Lan Yin[5], and Xiaomei Zhu[6]

1 *Department of Mechanical Engineering*, *University of Washington*, *Seattle*, *WA* 98195, *USA*
2 *Department of Physics*, *University of Washington*, *Seattle*, *WA* 98195, *USA*
3 *Department of Microbiology*, *University of Washington*, *Seattle*, *WA* 98195, *USA*
4 *Department of Chemical Engineering*, *University of Washington*, *Seattle*, *WA* 98195, *USA*
5 *School of Physics*, *Peking University*, 100871 *Beijing*, *PR China*
6 *GeneMath*, 5525 27th *Ave*. *N.E.*, *Seattle*, *WA* 98105, *USA*

**Abstract:** This is the first report on a systematic method for constructing a large scale kinetic metabolic model and its initial application to the modeling of central metabolism of *Methylobacterium extorquens* AM1, a methylotrophic and environmental important bacterium. Its central metabolic network includes formaldehyde metabolism, serine cycle, citric acid cycle, pentose phosphate pathway, gluconeogensis, PHB synthesis and acetyl-CoA conversion pathway, respiration and energy metabolism. Through a systematic and consistent procedure of finding a set of parameters in the physiological range we overcome an outstanding difficulty in large scale kinetic modeling: the requirement for a massive number of enzymatic reaction parameters. We are able to construct the kinetic model based on general biological considerations and incomplete experimental kinetic parameters. Our method consists of the following major steps: 1) using a generic enzymatic rate equation to reduce the number of enzymatic parameters to a minimum set while still preserving their characteristics; 2) using a set of steady state fluxes and metabolite concentrations in the physiological range as the expected output steady state fluxes and metabolite concentrations for the kinetic model to restrict the parametric space of enzymatic reactions; 3) choosing enzyme constants K's and K'$_{eq}$s optimized for reactions under physiological concentrations, if their experimental values are unknown; 4) for models which do not cover the entire metabolic network of the organisms, designing a dynamical exchange for the coupling between the metabolism represented in the model and the rest not included.
The success of our approach with incompletely input information is guaranteed by two known principles in biology, the robustness of the system and the cooperation among its various parts.

## Introduction

In the post genomic era, many of our current global concerns, such as health, energy, and environment, need perspectives from bioengineering. In response integrated and large scale wet and dry lab approaches associated with systems biology have been developing rapidly[1–3]. Microbiology has played and will continue to play an important role, because bacteria have several billion years of experience in exploring various living conditions. This approach may give us better ways to produce important medical agents, to increase the efficiency of using carbon sources, and to turn harmful materials into environmentally friendly ones. A key component in such endeavors is the mathematical modeling of large biological systems. Among them is metabolic modeling whose goal is to create a comprehensive multidimensional representation of all of the biosynthetic reactions in an organism. For this, we require mathematical models[4] to perform two ma jor functions. First, they should be able to reliably describe experimental observations[5], for instance, the models should indicate the metabolic fluxes and concentration of metabolites in each organelle or cell type at a given moment under specified conditions. Secondly, it is capable of generating experimentally[6,7] testable hypotheses leading to new experiments and new results which can then be used to refine the model. In the process of constructing such a model, new insights may often be gained as well. Modeling from the metabolic perspectives has been presented in many recent excellent reviews[8]. Metabolic networks have been discussed parallel to signaling networks[9], and the effect of noise in metabolic networks have also been discussed from various perspectives[10,11].

A survey of the literature suggests that models for metabolic networks are primarily (i) stoichiometric models[12] which display—in many cases on a genome-wide level—an organism's metabolic capabilities and (ii) kinetic models[13–15], which describe at the enzymatic level, the rate at which reactions proceed. Flux Balance Analysis (FBA)[16], in which mass conservation and other constraints are imposed on the metabolic network to determine a feasible solution space, is often used to evaluate the stoichiometric models. These constraints can be, for example, thermodynamic[17] and transcription regulatory[18]. Stoichiometric models can also be characterized by network-based pathway definitions[19]. While useful, important temporal behaviors are beyond the reach of FBA, such as the transient accumulation of toxic intermediate metabolites and the dynamical deficiency of an important nutrient. These time-dependent behaviors can drastically affect the life process of an organism. Nevertheless, kinetic models have not been studied as extensively as they should be. The common reasons are: a mechanistic formulation of even single enzyme kinetics is complicated with many parameters[20] and the experimental data for such parameters are scarce[7]. However, they are important for modeling behaviors such as oscillations[21,22] or bi-stability[23,24] that often occur in biological networks. Realistic kinetic modeling of large metabolic networks has been difficult. A major challenge is how to achieve biologically meaningful predictions from the mathematical model in the face of sparse experimental kinetic parameters and other necessary inputs.

In this paper, we present a systematic methodology to solve this important parameter issue. The first problem to be addressed is how to effectively and accurately represent an enzymatic reaction which may easily contain tens or more molecular parameters. Our solution is based on an observation that a complicated and exact (in the quasi-steady state sense) enzymatic rate equation can be rigorously cast into a generic form with the smallest set of kinetic parameters, which have transparent biochemical interpretations and can be directly related to experimental values.

The second problem to be addressed is that given the set of reactions how can a plausible set of fluxes be identified. We have at least two ways to solve this problem. One is to use a method related to flux balance analysis. Another is to set up a robust kinetic model which can generate various fluxes, although the kinetic parameters used may not be related to a realistic situation. The reason that realistic steady state fluxes can be obtained by crude kinetic models is that such fluxes are not sensitive to most kinetic parameters.

The third problem is how to obtain a consistent set of all parameters needed with given fluxes. Our solution is based on two considerations: to make use of as much available experimental data as we can, and to use a matching rule. Once a reasonable set of parameters is obtained, various predictions can be made, and, can be discussed in the biological contexts. In addition, various mathematical analyses can be carried to further test the consistency of parameters.

In order to make sure this scheme stays on the right track, we have been interacting closely with an on-going experimental study: We have been using the *Methylobacterium extorquens* AM1[25] as our model organism for validation and for inspiration. This bacterium is well studied as a facultative methylotroph[26], meaning it is able to utilize both single and multiple carbon unit compounds. A genome sequence[27] is available and a global analysis integrating data from the metabolome[28–30], transcriptome[31], proteome[32] and genome[25] is currently underway. The accumulation of multi-tier information makes *M. extorquens* AM1 a model system for integrative functional genomic analysis[28] and for quantifying changes between methylotrophic versus non-methylotrophic growth modes. Stoichiometric models have been constructed[33,34]. Furthermore, some key questions on the metabolism of C1 carbons and the glyoxylate regeneration cycle in *M. extorquens* AM1 is still under active investigation[35] making it an interesting as well as a challenging model organism.

## 1 Methods

### 1.1 Metabolic network of *M. extorquens* AM1

First, we will specify the large metabolic network we have been using for validation: the central metabolism of *M. extorquens* AM1(Fig. 1). The key pathways represented are formaldehyde metabolism, Serine cycle, Tricarboxylic acid (TCA) cycle, Pentose phosphate pathway (PPP), Poly-β-hydroxybutyrate (PHB) synthesis and Acetyl-CoA conversion pathway, gluconeogensis and serine biosynthesis[25].

## 1.2 Generic kinetic equations

There is a wide variety of kinetic rate equations depending on the type of binding mechanism and the type of modifiers. The large number of parameters makes it difficult, if at all possible, to determine all parameters experimentally. However, it is not always necessary to know the full mechanistic rate equation in order to correctly characterize the behavior of the organism. The reason is that physiologically, metabolite concentrations are usually restricted to a rather narrow subspace of the whole range of concentrations[36].

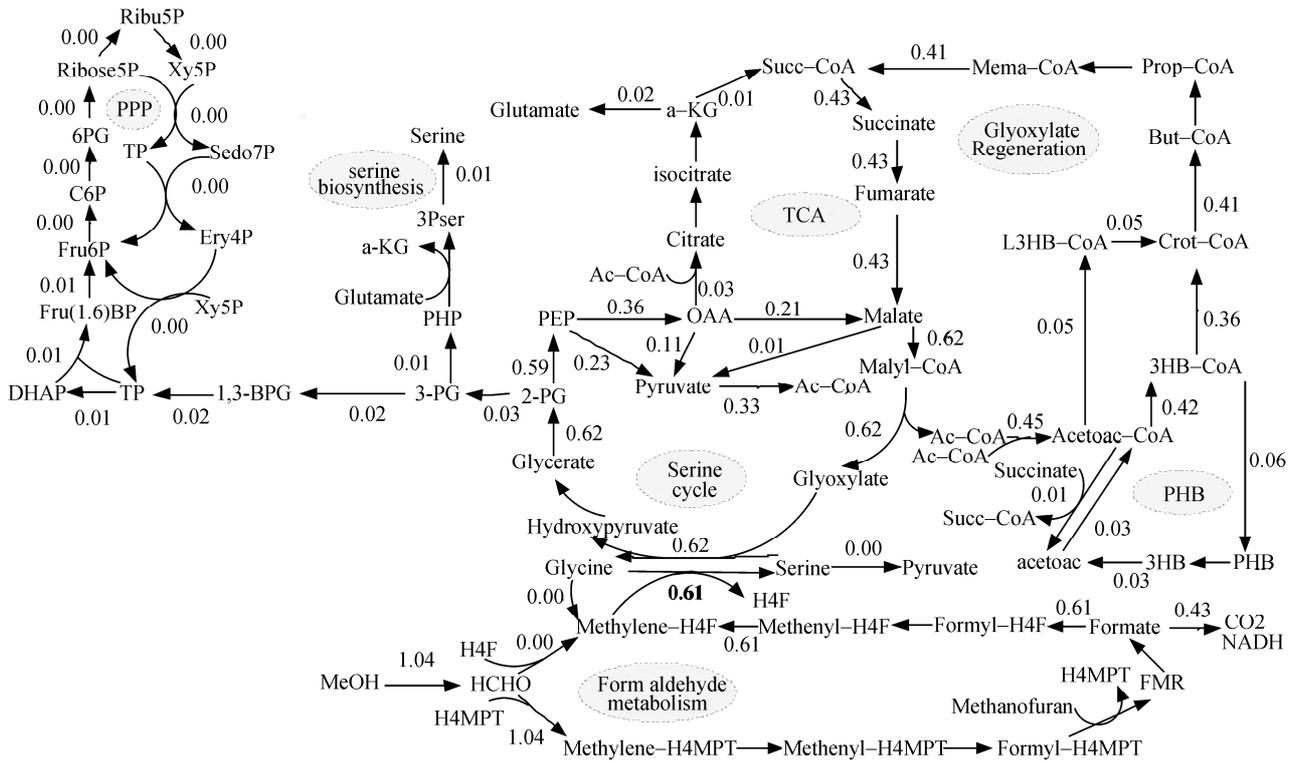

**Fig. 1** The main metabolic pathways of *Methylobacterium extorquens* AM1
The number with each reaction is its steady state flux calculated from the present kinetic model feeding on methanol (*c.f.* Table 3)

A chemical reaction can be written in the general form of Eq. (1).

$$A_1 + A_2 + ... + A_m \underset{V_B}{\overset{V_A}{\rightleftharpoons}} P_1 + P_2 + ... + P_n \quad (1)$$

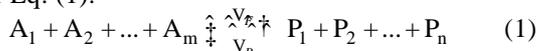

Each $A_i$ can be the same chemical species as the previous $A_i$ or they can be a different chemical species. The same holds for the products P. In this way, the stoichiometry is specified. Enzymes catalyze most biological reactions. Implicit in the above is an enzyme E which appears unbound on both sides of the reaction. The chemical reaction presented above does not indicate the sequence of reactions and the formation of intermediate compounds, such as enzyme- substrate complexes. The sequence of reactions leading to the enzyme-substrate complex is important when we seek to understand the catalytic properties of enzymes.

Even though various binding mechanisms leading to complicated rate equations have been studied in great detail[20], these mechanistic rate equations are of less practical use. This is because they contains a profusion of parameters even for a bi-substrate ordered mechanism that are difficult to determine, even for *in vitro* studies of enzymes. Another problem is that even though databases of enzyme data have been established, there is still a lack of data from enzyme characterization under physiologically relevant conditions, *i.e.* with appropriate pH, temperature and metabolite pools. Compounding the problem is a lack of standardized procedures for reporting results from enzyme studies. Various inconsistencies that arise from using enzyme databases have been summarized[36]. Efforts have been made towards more uniform data reporting[37].

To alleviate some of the problems we highlighted, we use a generic form for the rate equations[38] for the set of metabolic reactions in our model. This produces a minimal set of relevant parameters for characterizing enzymatic reactions. The generic enzymatic rate equation, Eq. (2) can be constructed from the general chemical equation in Eq. (1). Such equation has been motivated by various computational and biological considerations, a result of close interaction between the experimental and computational efforts.

$$u([A_i],[P_j]) = \frac{V_F \prod_{i=1}^{m} \frac{[A_i]}{K_i} - V_B \prod_{j=1}^{n} \frac{P_j}{K_j}}{f_1(V_F,V_B)\prod_{i=1}^{m}(1+\frac{[A_i]}{K_i}) + f_2(V_F,V_B)\prod_{j=1}^{n}(1+\frac{P_j}{K_j})} \quad (2)$$

The two functions $f_1$ and $f_2$ have the following properties:

$$f_1(V_F, V_B) + f_2(V_F, V_B) = 1 \quad (3)$$
$$f_1(V_F=0, V_B)=0 \quad (4)$$
$$f_2(V_F, V_B=0)=0 \quad (5)$$

The first property is a normalization condition so that at very low concentrations of $A$'s and $P$'s, the denominator is unity. The second property expresses the fact that if the reaction is only backward (i.e. $V_F$ is zero so that the first term in the numerator vanishes), the rate should not be affected by the concentration of $A$. Therefore, the first term in the denominator has to be zero also. The third property has the same requirement as the second property but applies for the case where the reaction is irreversible in the backward direction (i.e. $V_B$ is zero so that the second term in the numerator vanishes). The irreversibility is considered in the last elementary reaction where the enzyme product complex disassociates to give the free enzyme and product. We have shown that the full mechanistic rate equations can generally be recast into the proposed generic form[38].

### 1.3 Modifiers: inhibitors and activators

Modification of a reaction, or its control and regulation from an engineering perspective, can be in the form of substrate activators or product inhibitors. We discuss how they can be included in the generic rate equation. In enzyme kinetic studies, there are two common types of inhibitors: non-competitive and competitive[20]. Competitive inhibitors compete with the substrate for binding to the enzyme forming an enzyme-inhibitor complex. The inhibitor does not act on the enzyme-substrate complex once it is formed. The effects of inhibitor on the reaction flux can be written in the form

$$v \to \frac{v}{1+(I/a)^n}, \quad (6)$$

that is, the left hand of Eq.(2) times the inhibition factor, of sigmoidal form, where $I$ is the inhibiting product and $a$ is some numerical factor which controls how large $I$ has to be before the reaction rate is significantly reduced. The power of $I$ in the denominator (represented by $n$) can be modified depending on the extent of product inhibition[39] desired. As $n$ increases, the function becomes more step-like. We do not explicitly specify the type of inhibition for the inhibitor modifier that we use in Eq. (6). It is clear that we can account for non-competitive inhibitors effect since $V_F$ is changed by multiplying the modifier, same as in Eq. (6).

The activation can be handled similarly, that is, the resulting rate equation takes the form of

$$v \to v \frac{(X/b)^m}{1+(X/b)^m}, \quad (7)$$

the left hand of Eq.(2) times the activation factor of sigmoidal form, and $X$ is the activation product and $b$ some numerical factor which controls how large $X$ has to be before the reaction rate is significantly reduced and $m$ is an integer larger than 1.

To summarize, we use a generic rate equation for our reactions. This general equation requires a minimum set of parameters: the maximal forward and backward reaction velocity: $V_F$ and $V_B$ and $K_i$ as the apparent Michaelis-Menten parameter for each reactant, which defines how close each reactant is to saturation. The advantage of using a generic kinetic equation to construct metabolic models is obvious. Each reaction can be specified in the same way in a parameter file, which can be easily interpreted by a computer program. We do not have to worry about the different binding mechanism with different parameters for each separate equation.

### 1.4 Substrates and metabolites

The metabolic reaction network consists of 80 reactions and 80 metabolites. The list of metabolites and their corresponding abbreviations is shown in Table 1. A few substrates are assumed to be present in large excess so we ignored their effects on the reaction rates by setting their concentration to be constant. Apart from these considerations, we assume that all reactions take place in just one compartment within the cytoplasm and the reactions inside the cell are not restricted by the diffusion process.

Biomass production was represented as a sum of 20 key precursors[33]. This is shown in Table 1. Under realistic situations, biomass fluxes should depend on metabolite availability. If metabolite $A$ is extracted for biomass production (i.e. $b<0$), we model the biomass flux for $A$ as $2b[A]/([A]_0+[A])$ where $[A]_0$ is the expected steady state concentration of metabolite $A$. On the other hand, if metabolite $B$ comes from biomass production (i.e. $b>0$), we model the biomass flux for $B$ as $2b[B]_0/([B]_0+[B])$ where $[B]_0$ is the expected steady state concentration of metabolite $B$.

When the metabolite concentrations reaches the expected steady state, i.e. $[A]=[A]_0$ and $[B]=[B]_0$, we see that the biomass flux is exactly $b$.

**Table 1** List of metabolites in *Methylobacterium extorquens* AM1 central metabolism and steady state concentrations of kinetic model

| # | Metabolites | Abbreviation | Biomass composition, $b$ | Steady state concentration, $x_0$ (mmol/L) |
|---|---|---|---|---|
| 1 | Nicotinamide adenine dinucleotide | NAD | −5 | 1.69 |
| 2 | Formaldehyde | HCHO | 0 | 0.01 |
| 3 | 5,10-Methylene-tetrahydrofolate | Methylene-H4F | 0 | 0.07 |
| 4 | 5,10-Methenyl-tetrahydrofolate | Methenyl-H4F | 0 | 0.07 |
| 5 | Nicotinamide adenine dinucleotide phosphate | NADP | 257 | 0.62 |
| 6 | Formyl-tetrahydromethanopterin | Formyl-H4F | 0 | 0.07 |
| 7 | Formate | Formate | 0 | 2.00 |
| 8 | Adenosine triphosphate | ATP | −585 | 2.88 |
| 9 | 5,10-Methylene-tetrahydromethanopterin | Methylene-H4MPT | 0 | 0.20 |
| 10 | 5,10-Methenyl-tetrahydromethanopterin | Methenyl-H4MPT | 0 | 0.20 |
| 11 | Formyl-tetrahydromethanopterin | Formyl-H4MPT | 0 | 0.02 |
| 12 | Formylmethanofuran | FMR | 0 | 0.01 |
| 13 | Glycine | Gly | −13 | 16.58 |
| 14 | Serine | Ser | −7 | 4.90 |
| 15 | Glyoxylate | Glyox | 0 | 12.60 |
| 16 | Hydropyruvate | h-pyruvate | 0 | 1.40 |
| 17 | Glycerate | Glycerate | 0 | 2.22 |
| 18 | 3-phosphoglycerate | 3-PG | 0 | 2.36 |
| 19 | 2-phosphoglycerate | 2-PG | 0 | 2.19 |
| 20 | Phosphoenolpyruvate | PEP | −11 | 6.00 |
| 21 | Oxaloacetate | OAA | −41 | 7.35 |
| 22 | Malate | Malate | 0 | 1.60 |
| 23 | Malyl-CoA | Malyl-CoA | 0 | 3.43 |
| 24 | Acetyl-CoA | Acetyl-CoA | −53 | 0.38 |
| 25 | Pyruvate | Pyr | −42 | 3.85 |
| 26 | Coenzyme A | CoA | 60 | 0.10 |
| 27 | Citrate | Cit | 0 | 4.74 |
| 28 | Cis-aconitate | Cis-acon | 0 | 1.00 |
| 29 | Isocitrate | Iso-C | 0 | 1.00 |
| 30 | Alpha-Ketoglutarate | α-KG | −17 | 0.65 |
| 31 | Succinyl-CoA | Succ-CoA | −7 | 0.65 |
| 32 | Guanosine diphosphate | GDP | 0 | 0.11 |
| 33 | Succinate | Succ | 7 | 11.77 |
| 34 | Fumarate | Fum | 0 | 0.51 |
| 35 | Flavin adenine dinucleotide | FAD | 0 | 0.99 |
| 36 | Glucose 6-phosphate | Glc6P | 0 | 4.20 |
| 37 | Fructose 6-phosphate | Fru6P | −16 | 6.36 |
| 38 | Fructose 1,6-biphosphate | Fru (1,6)-BP | 0 | 2.00 |
| 39 | Dihydroxyacetone phosphate | DHAP | 0 | 12.36 |
| 40 | Glyceraldehyde 3-phosphate | TP | −2 | 12.41 |
| 41 | 1,3-biphosphoglycerate | 1,3-BPG | 0 | 11.31 |
| 42 | 6-phosphogluconate | 6-PG | 0 | 2.00 |
| 43 | Ribose 5-phosphate | Ribose5P | −10 | 4.51 |
| 44 | Ribulose 5-phosphate | Ribu5P | 0 | 4.51 |

| # | Metabolites | Abbreviation | Biomass composition, $b$ | Steady state concentration, $x_0$ (mmol/L) |
|---|---|---|---|---|
| 45 | Xylulose 5-phosphate | Xy5P | 0 | 1.80 |
| 46 | Sedoheptolose-7-phosphate | Sedo7P | 0 | 1.85 |
| 47 | Erythrose-4-phosphate | Ery4P | −5 | 3.59 |
| 48 | Acetoacetyl-CoA | Acetoac-CoA | 0 | 0.18 |
| 49 | β-hydroxybutyrate-CoA | 3HB-CoA | 0 | 0.03 |
| 50 | Poly-β-hydroxybutyrate | PHB | −93 | 0.12 |
| 51 | β-hydroxybutyrate | 3HB | 0 | 1.16 |
| 52 | Acetoacetate | Acetoac | 0 | 0.85 |
| 53 | Crotonyl-CoA | Crot-CoA | 0 | 0.01 |
| 54 | (L)-β-hydroxybutyryl-CoA | L3HB-CoA | 0 | 0.24 |
| 55 | Butyryl-CoA | But-CoA | 0 | 0.22 |
| 56 | Propionyl-CoA | Prop-CoA | 0 | 0.09 |
| 57 | Methylmalonyl-CoA | Mema-Coa | 0 | 0.27 |
| 58 | 3-phosphohydroxypyruvate | PHP | 0 | 0.60 |
| 59 | 3-Phosphoserine | 3Pser | 0 | 0.09 |
| 60 | Glutamate | Glu | −17 | 12.00 |
| 61 | Ubiquinone | Q | 0 | 0.98 |
| 62 | Cytochrome-c (oxidized) | Cyt-Cox | 0 | 1.00 |
| 63 | Nicotinamide adenine dinucleotide (reduced) | NADH | 5 | 0.53 |
| 64 | Nicotinamide adenine dinucleotide phosphate (reduced) | NADPH | −257 | 0.42 |
| 65 | Adenosine diphosphate | ADP | 585 | 0.60 |
| 66 | Guanosine triphosphate | GTP | 0 | 0.99 |
| 67 | Flavin adenine dinucleotide (reduced) | FADH2 | 0 | 0.11 |
| 68 | Ubiquinol | QH2 | 0 | 1.02 |
| 69 | Cytochrome-c (reduced) | Cyt-Cred | 0 | 1.00 |
| 70 | Methanol | MeOH | 0 | 100.00 |
| 71 | Tetrahydrofolate | H4F | 0 | 0.07 |
| 72 | Tetrahydromethanopterin | H4MPT | 0 | 0.20 |
| 73 | Carbon dioxide | CO2 | 0 | 5.00 |
| 74 | Ammonia | NH3 | 0 | 1.00 |
| 75 | Methanofuran | MFR | 0 | 0.40 |
| 76 | Phosphate | Pi | 0 | 3.00 |
| 77 | Proton | H | 0 | 6.31 |
| 78 | External Pyruvate | Py-Out | 0 | N/A |
| 79 | External Succinate | Succ-Out | 0 | N/A |
| 80 | Electron | e | 0 | 0.00 |

MeOH, H$_4$F, H$_4$MPT, methanofuran, CO$_2$, NH$_3$, Pi, CoA, Cyt-cox, Cyt-cred, pyruvate-uut, succinate-out are constants. Another set of constants (in mmol/L) are NAD + NADH=2.22, NADP + NADPH =1.04, ATP + ADP =3.48, GDP + GTP =1.1, FAD + FADH2=1.1, Q + QH2 = 2.0

The biomass flux is also flexible in that if more of metabolite [$A$] (relative to the expected steady state level) is available, a larger portion is extracted for biomass production. A lesser portion is extracted when the concentration of $A$ falls below the expected steady state level.

**1.5 Incorporation of fluxes and metabolite concentrations a priori to determine parameters for enzymatic rate equations**

We start by analyzing the fluxes obtained from enzymatic rate equations. For the purpose of demonstration, consider a reversible reaction among two metabolites, A ↔ B:

$$v = \frac{V_F \frac{A}{K_A} - V_B \frac{B}{K_B}}{f_1(V_F, V_B)\left(1 + \frac{[A]}{K_A}\right) + f_2(V_F, V_B)\left(1 + \frac{[B]}{K_B}\right)}, \quad (8)$$

here

$$f_1=V_F^2/(V_F^2+V_B^2), f_2= V_B^2/(V_F^2+V_B^2)$$

If we require the flux $v$ to be the expected steady state value when the metabolite concentrations are the expected steady state values, the above equation constrains the allowed choices of $V_F$, $V_B$, $K_A$, $K_B$. If experimental values of $K_A$, $K_B$, $K_{eq} = (V_F/K_A)/(V_B/K_B)$, are known, then the above equation allows the determination of $V_F$. If $K_A$, $K_B$, $K_{eq}$ are not measured, according to biochemical observation and the optimization principle we presented, later, we propose to choose $K_A$, $K_B$ in the range of steady state metabolite concentrations. For simplicity, we choose $K_A=[A]_0$, $K_B=[B]_0$, where $[A]_0$ and $[B]_0$ are desired steady state concentrations. Our optimization analysis also demonstrates that the equilibrium constant $K_{eq} = (V_F/K_A)/(V_B/K_B)$ is optimized on the order of $[B]_0/[A]_0$. When considering thermodynamic constrains on $K_{eq}$, such an optimization could be interpreted as optimization of the metabolite concentrations. For simplicity we choose

$$K_{eq} = 2\,[B]_0/[A]_0$$

for forward reactions, $v_0 > 0$ and

$$K_{eq} = 0.5\,[B]_0/[A]_0$$

for backward reactions, $v_0 < 0$.

By using these $K_A$, $K_B$, $K_{eq}$, $V_F$ are determined.

In summary, we demonstrate that how parameters in the enzymatic rate equations $V_F$, $V_B$, $K_A$, $K_B$ are determined by requiring steady state fluxes and steady state metabolites concentrations, together with the requirement that $K_A$, $K_B$ and $K_{eq}$ take values in the range of metabolite concentrations. Reactions involving more metabolites with inhibitors and activators can be determined in a similar manner.

### 1.6 Optimization of $K$'s and $K'$eqs based on fluctuation minimization

First consider an irreversible reaction $A\rightarrow$. The generic rate equation is given by

$$v = \frac{V_F[A]/K_A}{(1+[A]/K_A)} \quad (9)$$

We calculate how the enzymatic rate changes with disturbances of metabolite concentration. If metabolite concentration deviates by $\delta[A]$, then the relative change of the flux is given by $\delta\ln v = (\partial\ln v/\partial\ln[A])\,\delta\ln[A]$. If, on the other hand, flux changes by an amount $\delta v$, for instance due to the change of reaction which produces metabolite A, the amount of relative change in metabolite is given by $\delta\ln[A]= (\partial\ln[A]/\partial\ln v)\delta\ln v$. For the stability of the metabolic network, both these changes should be kept small. Since the coefficients $(\partial\ln v/\partial\ln[A])$ and $(\partial\ln[A]/\partial\ln v)$ are reciprocal to each other, compromises need to be made. Carrying out calculations we have

$$\frac{\partial \ln v}{\partial \ln[A]} = \frac{K_A}{K_A+A}, \quad (10)$$

which changes from 0 to 1 with $K_A$ increasing from 0 to $\infty$, and

$$\frac{\partial \ln[A]}{\partial \ln v} = \frac{K_A+A}{K_A}, \quad (11)$$

which changes from $\infty$ to 1 with increasing $K_A$ from 0 to $\infty$. The choice of $K_A \sim [A]$ is apparently a good compromise because it makes both $(\partial\ln v/\partial\ln[A])$ and $(\partial\ln[A]/\partial\ln v)$ reasonably small. For simplicity we choose $K_A = [A]_0$, where $[A]_0$ is the expected steady state value. With this choice, the steady state flux is $v_0 = V_F/2$. Therefore for consistency, $V_F$ should be chosen as $2v_0$.

Next we consider a reversible reaction

$$v = \frac{V_F\dfrac{[A]}{K_A}-V_B\dfrac{B}{K_B}}{f_1(V_F,V_B)\left(1+\dfrac{[A]}{K_A}\right)+f_2(V_F,V_B)\left(1+\dfrac{[B]}{K_B}\right)}, \quad (12)$$

With

$$f_1=V_F^2/(V_F^2+V_B^2),\ f_2=V_B^2/(V_F^2+V_B^2)$$

and $v>0$.

We present an intuitive approach to analyze such a reaction. We assume the reaction as forward ($v>0$) and for approximation, we ignore the backward reaction when choosing the forward reaction constants $K_A$ and $V_F$. This gives $V_f=2v_0$ and $K_A=[A]_0$, according to the results presented above for the case of $A\rightarrow$. Considering that metabolite A is at its steady state value $[A] = [A]_0$ so that the focus of the rate equation is only on $[B]$, we have a simplified relationship

$$v = \frac{2v_0 - V_B\dfrac{[B]}{K_B}}{2f_1(V_0,V_B)+f_2(2v_0,V_B)\left(1+\dfrac{[B]}{K_B}\right)}, \quad (13)$$

We now require the absolute values of both $(\partial\ln v/\partial\ln[B])$ and $(\partial\ln[B]/\partial\ln v)$ to be small to avoid large fluctuations due to $\delta\ln[B]$ and $\delta\ln v$. Carrying out the calculation we have

$$\frac{\partial \ln v}{\partial \ln[B]} = \frac{-[B]((2f_1+f_2)V_B+2f_2v_0)}{K_B^v\left[2f_1+f_2\left(1+\dfrac{[B]}{K_B}\right)\right]^2}, \quad (14)$$

and $\partial\ln[B]/\partial\ln v = 1/(\partial\ln v/\partial\ln[B])$.

The values of $\partial\ln[B]/\partial\ln v$ are $-\infty$ both when $K_B=0$ and $K_B=\infty$. Apparently some physiological value of $K_B$ between 0 and $\infty$ is needed to avoid large fluctuations due to $\partial\ln[B]/\partial\ln v$. A choice of $\partial\ln[B]/\partial\ln v = -20/11\approx-1.82$ will satisfy such a requirement. There are positive solutions for $K_B$ for any given $V_B$, when $V_B$ is in a wide range of values, since $\partial\ln[B]/\partial\ln v = -20/11$ is a polynomial equation of second order for $K_B$ if $V_B$ is treated as a constant. For example if $V_B = v_0$, then $K_B=[B]_0$ is a solution for $\partial\ln[B]/\partial\ln v =-20/11$. In such a case large fluctuations of $\delta\ln[B]=(\partial\ln[B]/\partial\ln\nu)\,\delta\ln\nu$ and $\delta\nu=(\partial\ln\nu/\partial\ln[B])\,\delta\ln[B]$ are all avoided.

Next we need to find if there is any reason to choose $V_B=\nu_0$ from optimization of enzyme functions. For the purpose of examining the enzyme performance due to variation in $V_B$, we let $K_A=[A]_0$ $K_B=[B]_0$, then at the steady state, the flux is simply given by $\nu = (V_F-V_B)/2$. The change of $\nu$ due to change of $V_B$ is $\delta\ln\nu= (\partial\ln\nu/\partial\ln V_B)\,\delta\ln V_B$, with $(\partial\ln\nu/\partial\ln V_B)=-V_B/2\nu_0$, assuming $\nu=\nu_0$. If $V_B$ fluctuates, for instance due to pH values, then $\nu$ changes accordingly.

If both $V_F$ and $V_B$ are very large compared with $\nu_0$ so that $\nu_0 = (V_F-V_B)/2$ is the result of large forward and backward reactions subtracting each other, for instance $V_F=1002\nu_0$ and $V_B=1000\nu_0$, $\nu_0 = (V_F-V_B)/2 = (10002\nu_0-1000\nu_0)/2$, the fluctuation of $\nu$ due to fluctuation of $V_B$ is large because $\delta\ln\nu= (\partial\ln\nu/\partial\ln V_B)\,\delta\ln V_B =-(V_B/2\nu_0)\delta\ln V_B =-500\,\delta\ln V_B$. Such an enzyme is not optimized with respect to fluctuations of $V_B$. On the other hand, $V_B$ should not be too small because if $V_B=0$, then this reaction is essentially an irreversible one, resulting in a divergent equilibrium constant $K_{eq}=(V_F/K_A)/(V_B/K_B)$, inconsistent with the assumption of a reversible reaction.

With these complex relationships concerning fluctuation and thermodynamic restriction of equilibrium constant, a consistent choice for the enzyme is to have $V_B$ on the order of physiological flux $\nu_0$, which also needs $V_F$ on the order of $\nu_0$ because of the relationship $\nu_0 = (V_F-V_B)/2$. In such a way, fluctuations due to the change of $V_B$ are small while the equilibrium constant $K_{eq} = (V_F/K_A)/(V_B/K_B)$ can be maintained at the desired value of $K_{eq} = (V_F/[A]_0)/(V_B/[B]_0)$ by adjusting steady state metabolite concentrations $[A]_0$, $[B]_0$.

Finally, the above discussion can be formalized by designing a cost function to minimize fluctuations in fluxes due to relative changes in metabolite concentrations $[A]$ and $[B]$, in $V_f$ and $V_B$, and fluctuations of metabolite concentrations due to changes in fluxes simultaneously. We have performed such an optimization process. Since it is only a formalization of the analysis presented in this section, the qualitative results are the same. We will not present such an approach here.

To summarize, we have demonstrated that the choices of $K_A$, $K_B$, $K_{eq}$, in the physiological range of $[A]_0$, $[B]_0$ and $[B]_0/[A]_0$ are optimized for minimizing fluctuations. For simplicity we use $K_A=[A]_0$, $K_B=[B]_0$, and $K_{eq}=2[B]_0/[A]_0$ in our model. For reactions involving more metabolites, similar analysis can be carried through. We found that the results are similar, which is reasonable because the generic rate equation has the same form. The principle behind the approach in this section is essentially a matching condition similar to the Hebbian learning rule in neurodynamics[40].

## 1.7 Coupling to the rest of the metabolic network not included in the model

In many cases, we wish to model a particular section of the metabolic network instead of the whole network in the organism. For example, most of the experimental works on *M. extorquens* AM1 have been concentrated on central metabolism of this bacterium[25]. Therefore a kinetic model of *M. extorquens* AM1 central metabolism is useful at this stage. When modeling only part of the metabolic network, its coupling to the rest of metabolic network not included in the model should contain both the steady state values of fluxes in and out of the metabolic system in the model and their dynamics.

First consider a situation when two metabolites, A within the model and B not, are coupled with a reversible reaction A↔B with $\nu>0$. Since B is not included in the model, this reaction is not included either. Instead we consider that a flux $f_A$ is taking out A, e.g. for biomass synthesis, in the kinetic equation of metabolite $[A]$ by subtracting a term $\eta f_A[A]/(k + [A])$, where $\eta= (k + [A]_0)/[A]_0$, $[A]_0$ the expected steady state value of $[A]$. Such a subtraction is more realistic than simply subtracting $f_A$, because if metabolite B were included in the model, there would be term representing A↔B in the kinetic equation of metabolite $[A]$, which is a subtraction of similar form. Therefore we propose when there is a flux taken out of the system, this form should be used for coupling. For simplicity we choose k= $[A]_0$ and $\eta=2$.

Next consider for metabolites A and B connected with a reaction A↔B with $\nu>0$, and B is within the model while A is not. At steady state there is a constant input of flux from A to B into the system in the model. Now consider the rate equation for A↔B, the flux added in to B is in the form of $(\mu - \nu[B])/f([B])$, with $\mu$ and $\nu$ constants and $f([B])$ a function of $[B]$. A form to mimic such coupling is a function $\eta' f_B/(k' + [B])$, with $\eta'= k' + [B]_0$, $[B]_0$ the expected steady state value of $[B]$, $f_B$ the flux adding to B. Again for simplicity we choose k′= $[B]_0$ and $\eta'=2[B]_0$.

To summarize, if a flux $f_A$ is taking out from metabolite A to be used beyond the modeled system, we subtract a term $2f_A[A]/([A]_0 + [A])$ from the dynamic equation of $[A]$. If a flux $f_B$ is adding to metabolite B from sources outside the modeled system, we add a term $2f_B[B]_0/([B]_0 + [B])$ to the dynamic equation of $[B]$.

## 1.8 Central Metabolism of *M. extorquens* AM1

Methylotrophic bacterium *M. extorquens* AM1 is a well studied facultative methylotroph[25,33] which utilizes both single and multiple carbon unit compounds. The central metabolism adjusts to growth in different conditions by changing expression of enzymes under different growth conditions[25,33]. The metabolic pathways necessary for energy balance and growth need for carbon under methylotropic growth conditions are included in our model, including formaldehyde metabolism[41], the serine cycle[42], the citric acid cycle, the pentose phosphate pathway, poly-β-hydroxybutyrate (PHB) synthesis, the gloxylate regeneration cycle[43,44], gluconeogenesis, serine biosynthesis and respiratory chain.

Under methylotropic growth conditions, the serine cycle is utilized for $C_1$ assimilation as follows: methanol (MeOH), a $C_1$ compound is first oxidized to formaldehyde by the methanol dehydrogenase complex in the periplasm. The formaldehyde that enters the cytoplasm condenses with either tetrahydrofolate ($H_4F$) or tetrahydromethanopterin ($H_4MPT$) to form the respective methylene derivatives. The reaction of formaldehyde with $H_4F$ to produce methylene-$H_4F$ is considered a spontaneous reaction since no enzyme that catalyzes this reaction has been found thus far. Methylene-$H_4F$ is either assimilated through the serine cycle or oxidized to methenyl-$H_4F$, formyl-$H_4F$ and eventually to formate. Formate is further oxidized to $CO_2$ by formate dehydrogenases in an energy-generating reaction. Alternatively, formaldehyde is oxidized via the $H_4MPT$-linked pathway. In the initial step for this pathway, formaldehyde reacts with $H_4MPT$ to form Methylene-$H_4MPT$.

*M. extorquens* AM1 has a complete citric acid cycle since it can grow on alternative carbon source such as pyruvate. The citric acid cycle is dispensable under $C_1$ growth condition as indicated by mutants that lack α-KG dehydrogenase activity[45]. The citric acid cycle overlaps in part with the gloxylate regeneration cycle, which converts acetyl-CoA to glyoxylate to maintain the serine cycle. The glyoxylate regeneration cycle also overlaps with PHB synthesis.

## 2 Results and discussions

### 2.1 Kinetic equations for metabolic network

After going though the procedure described above, a set of dynamical equations corresponding to the reactions of the metabolic network can be explicitly obtained, which are in the form of differential equations and may be written as

$$\frac{dx}{dt} = Sv(x) + b(x) \quad (15)$$

where $x(t)$ is a vector of metabolite concentration at time $t$ with dimensions 80×1. $S$ is the stoichiometric matrix with dimensions 80×80. $v$ is a vector (dimensions 80×1) containing the reaction rate or fluxes and is a function of $x(t)$, the metabolite concentration at time $t$. The vector $b(x)$ is the flux for biomass production as described in Sec. 2.4. $x$ represents the state of the system. The differential equations are tabulated in Table 2 in the stoichiometric form, and the kinetic parameters are given in Table 3 contains the references[46–53].

Starting from an initial condition $x_0$, the set of differential equations is integrated numerically in Matlab using the ordinary differential solver, ode15s until we reach a steady state solution, $x(t_{final})$. The set of steady state fluxes can be obtained by substituting the final metabolite concentrations, $x$ into $v$. We present our results from the kinetic simulation in the next section.

With the explicitly mathematical representation of the metabolic network, various predictions can be obtained via standard mathematical tools which are usually available in Matlab or MATHEMATICA. An initial set of an exploration will be given below to show the feasibility of our present method.

### 2.2 Fluxes and concentrations

From our kinetic simulation, the flux distribution and metabolites concentrations can be predicted as a function of time. The predicted steady state flux distribution when growing on methanol is shown in Fig. 1 which can be obtained by methods based on FBA. The predicted steady state concentrations are in Table 1. The set of fluxes is shown in Table 3. The predicted TCA cycle fluxes are smaller compared to the predicted fluxes in the glyoxylate cycle. Some fluxes of central carbon metabolism with methanol as a substrate have been measured[54]. Low fluxes were detected through pyruvate dehydrogenase (Reaction No. 24, Table 3), α-KG dehydrogenase (Reaction no. 29, Table 3) and malic enzyme (Reaction No. 42, Table 3) for the wild-type *M. extorquens* AM1 growing on methanol.

**Table 2** Kinetic Equations for *Methylobacterium extorquens* AM1 central metabolic network in explicit stoichiometric form

| Variable | Metabolite | Kinetic equation | Variable | Metabolite | Kinetic equation |
| --- | --- | --- | --- | --- | --- |
| $x_1$ | NAD | $dx_1/dt = -v_6 - v_8 + v_{15} + v_{21} - v_{24} - v_{29} - v_{37} - v_{42} - v_{54} - v_{58} - v_{60} - v_{63} + v_{67} + v_{73} - v_{78} + b_1$ | $x_{41}$ | 1,3-BPG | $dx_{41}/dt = v_{37} + v_{38} + b_{41}$ |
| $x_2$ | HCHO | $dx_2/dt = v_1 - v_2 - v_7 + b_2$ | $x_{42}$ | 6-PG | $dx_{42}/dt = v_{43} - v_{44} + b_{42}$ |
| $x_3$ | Methylene-H4F | $dx_3/dt = v_2 + v_3 - v_{13} + v_{78} + b_3$ | $x_{43}$ | Ribose5P | $dx_{43}/dt = v_{44} - v_{45} - v_{47} + b_{43}$ |

| | | | |  | | | |
|---|---|---|---|---|---|---|---|
| $x_4$ | Methenyl-H4F | $dx_4/dt = -v_3 - v_4 + b_4$ | | $x_{44}$ | Ribu5P | $dx_{44}/dt = v_{45} - v_{46} + b_{44}$ |
| $x_5$ | NADP | $dx_5/dt = v_3 - v_9 + v_{16} - v_{28} - v_{43} - v_{44} + v_{51} + v_{59} - v_{66} + v_{72} - v_{73} + b_5$ | | $x_{45}$ | Xy5P | $dx_{45}/dt = v_{46} - v_{47} - v_{49} + b_{45}$ |
| $x_6$ | Formyl-H4F | $dx_6/dt = v_4 + v_5 + b_6$ | | $x_{46}$ | Sedo7P | $dx_{46}/dt = v_{47} - v_{48} + b_{46}$ |
| $x_7$ | Formate | $dx_7/dt = -v_5 - v_6 + v_{12} + b_7$ | | $x_{47}$ | Ery4P | $dx_{47}/dt = v_{48} - v_{49} + b_{47}$ |
| $x_8$ | ATP | $dx_8/dt = -v_5 - v_{17} - v_{22} - v_{38} - v_{40} + v_{41} - v_{61} - v_{70} - v_{71} - v_{77} + b_8$ | | $x_{48}$ | Acetoac-CoA | $dx_{48}/dt = v_{50} - v_{51} + v_{55} + v_{58} + v_{77} + b_{48}$ |
| $x_9$ | Methylene-H4MPT | $dx_9/dt = v_7 - v_8 - v_9 + b_9$ | | $x_{49}$ | 3HB-CoA | $dx_{49}/dt = v_{51} - v_{52} + v_{56} + b_{49}$ |
| $x_{10}$ | Methenyl-H4MPT | $dx_{10}/dt = v_8 + v_9 - v_{10} + b_{10}$ | | $x_{50}$ | PHB | $dx_{50}/dt = v_{52} - v_{53} + b_{50}$ |
| $x_{11}$ | Formyl-H4MPT | $dx_{11}/dt = v_{10} - v_{11} + b_{11}$ | | $x_{51}$ | 3HB | $dx_{51}/dt = v_{53} - v_{54} + b_{51}$ |
| $x_{12}$ | FMR | $dx_{12}/dt = v_{11} - v_{12} + b_{12}$ | | $x_{52}$ | Acetoac | $dx_{52}/dt = v_{54} - v_{55} - v_{77} + b_{52}$ |
| $x_{13}$ | Gly | $dx_{13}/dt = -v_{13} + v_{14} - v_{78} + b_{13}$ | | $x_{53}$ | Crot-CoA | $dx_{53}/dt = -v_{56} + v_{57} - v_{59} + b_{53}$ |
| $x_{14}$ | Ser | $dx_{14}/dt = v_{13} - v_{14} + v_{65} - v_{76} + b_{14}$ | | $x_{54}$ | L3HB-CoA | $dx_{54}/dt = -v_{57} - v_{58} + b_{54}$ |
| $x_{15}$ | Glyox | $dx_{15}/dt = -v_{14} - v_{23} - 2v_{75} + b_{15}$ | | $x_{55}$ | But-CoA | $dx_{55}/dt = v_{59} - v_{60} + b_{55}$ |
| $x_{16}$ | h-pyruvate | $dx_{16}/dt = v_{14} - v_{15} - v_{16} + v_{75} + b_{16}$ | | $x_{56}$ | Prop-CoA | $dx_{56}/dt = v_{60} - v_{61} + b_{56}$ |
| $x_{17}$ | Glycerate | $dx_{17}/dt = v_{15} + v_{16} - v_{17} + b_{17}$ | | $x_{57}$ | Mema-Coa | $dx_{57}/dt = v_{61} - v_{62} + b_{57}$ |
| $x_{18}$ | 3-PG | $dx_{18}/dt = v_{18} - v_{38} - v_{63} + b_{18}$ | | $x_{58}$ | PHP | $dx_{58}/dt = v_{63} - v_{64} + b_{58}$ |
| $x_{19}$ | 2-PG | $dx_{19}/dt = v_{17} - v_{18} - v_{19} + b_{19}$ | | $x_{59}$ | 3Pser | $dx_{59}/dt = v_{64} - v_{65} + b_{59}$ |
| $x_{20}$ | PEP | $dx_{20}/dt = v_{19} - v_{20} + v_{39} - v_{41} + b_{20}$ | | $x_{60}$ | Glu | $dx_{60}/dt = -v_{64} - v_{66} + b_{60}$ |
| $x_{21}$ | OAA | $dx_{21}/dt = v_{20} - v_{21} - v_{25} - v_{39} + v_{40} + b_{21}$ | | $x_{61}$ | Q | $dx_{61}/dt = -v_{67} + v_{68} - v_{69} - v_{72} - v_{74} + b_{61}$ |
| $x_{22}$ | Malate | $dx_{22}/dt = v_{21} - v_{22} + v_{32} - v_{42} + b_{22}$ | | $x_{62}$ | Cyt-Cox | $dx_{62}/dt = 0$ |
| $x_{23}$ | Malyl-CoA | $dx_{23}/dt = v_{22} + v_{23} + b_{23}$ | | $x_{63}$ | NADH | $dx_{63}/dt = v_6 + v_8 - v_{15} - v_{21} + v_{24} + v_{29} + v_{37} + v_{42} + v_{54} + v_{58} + v_{60} + v_{63} - v_{67} - v_{73} + v_{78} + b_{63}$ |
| $x_{24}$ | Acetyl-CoA | $dx_{24}/dt = -v_{23} + v_{24} - v_{25} - 2v_{50} + b_{24}$ | | $x_{64}$ | NADPH | $dx_{64}/dt = -v_3 + v_9 + v_{16} + v_{28} + v_{43} + v_{44} - v_{51} - v_{59} + v_{66} - v_{72} - v_{73} + b_{64}$ |
| $x_{25}$ | Pyr | $dx_{25}/dt = -v_{24} - v_{40} + v_{41} + v_{42} + v_{76} + v_{79} + b_{25}$ | | $x_{65}$ | ADP | $dx_{65}/dt = v_5 + v_{17} + v_{22} + v_{38} + v_{40} - v_{41} + v_{61} + v_{70} + v_{71} + v_{77} + b_{65}$ |
| $x_{26}$ | CoA | $dx_{26}/dt = 0$ | | $x_{66}$ | GTP | $dx_{66}/dt = -v_{30} - v_{39} + v_{71} + b_{66}$ |
| $x_{27}$ | Cit | $dx_{27}/dt = v_{25} + v_{26} + b_{27}$ | | $x_{67}$ | FADH2 | $dx_{67}/dt = -v_{31} + v_{60} - v_{69} + b_{67}$ |
| $x_{28}$ | Cis-acon | $dx_{28}/dt = -v_{26} + v_{27} + b_{28}$ | | $x_{68}$ | QH2 | $dx_{68}/dt = v_{67} - v_{68} + v_{69} + v_{72} + v_{74} + b_{68}$ |
| $x_{29}$ | Iso-C | $dx_{29}/dt = -v_{27} - v_{28} + b_{29}$ | | $x_{69}$ | Cyt-Cred | $dx_{69}/dt = 0$ |
| $x_{30}$ | α-KG | $dx_{30}/dt = v_{28} - v_{29} + v_{64} + v_{66} + b_{30}$ | | $x_{70}$ | MeOH | $dx_{70}/dt = 0$ |
| $x_{31}$ | Succ-CoA | $dx_{31}/dt = v_{29} + v_{30} - v_{55} + v_{62} + b_{31}$ | | $x_{71}$ | H4F | $dx_{71}/dt = 0$ |
| $x_{32}$ | GDP | $dx_{32}/dt = v_{30} + v_{39} - v_{71} + b_{32}$ | | $x_{72}$ | H4MPT | $dx_{72}/dt = 0$ |
| $x_{33}$ | Succ | $dx_{33}/dt = -v_{30} + v_{31} + v_{55} + v_{80} + b_{33}$ | | $x_{73}$ | CO2 | $dx_{73}/dt = 0$ |
| $x_{34}$ | Fum | $dx_{34}/dt = -v_{31} - v_{32} + b_{34}$ | | $x_{74}$ | NH3 | $dx_{74}/dt = 0$ |
| $x_{35}$ | FAD | $dx_{35}/dt = v_{31} + v_{60} + v_{69} + b_{35}$ | | $x_{75}$ | MFR | $dx_{75}/dt = 0$ |
| $x_{36}$ | Glc6P | $dx_{36}/dt = -v_{33} - v_{43} + b_{36}$ | | $x_{76}$ | Pi | $dx_{76}/dt = 0$ |
| $x_{37}$ | Fru6P | $dx_{37}/dt = v_{33} + v_{34} + v_{48} + v_{49} + b_{37}$ | | $x_{77}$ | H | $dx_{77}/dt = -v_{67} + 2v_{78} + 2v_{70} - v_{72} + b_{77}$ |
| $x_{38}$ | Fru(1,6)-BP | $dx_{38}/dt = -v_{34} - v_{35} + b_{38}$ | | $x_{78}$ | Py-Out | $dx_{78}/dt = 0$ |
| $x_{39}$ | DHAP | $dx_{39}/dt = v_{35} + v_{36} + b_{39}$ | | $x_{79}$ | Succ-Out | $dx_{79}/dt = 0$ |
| $x_{40}$ | TP | $dx_{40}/dt = v_{35} - v_{36} - v_{37} + v_{47} - v_{48} + v_{49} + b_{40}$ | | $x_{80}$ | E | $dx_{80}/dt = 2v_1 - 2v_{74} + b_{80}$ |

Vector $x$ stands for the metabolite concentrations, i.e., $x_3$ = [Methylene–H4F], $x_5$ = [NADP], $x_{48}$ = [Acetoac–CoA]

**Table 3** Kinetic parameters for the dynamical model of AM1 central metabolic network and the steady state fluxes

| Reactions | $V_F$ | $V_B$ | $K_1$ | $K_2$ | $K_3$ | $K_1'$ | $K_2'$ | $K_3'$ | $i_1$ | $i_2$ | $i_3$ | $j_1'$ | $j_2'$ | $j_3'$ | s.s.f |
|---|---|---|---|---|---|---|---|---|---|---|---|---|---|---|---|
| 1 MeOH→HCHO+2e | 2.97 | 0 | 10 | 0 | 0 | 0 | 0 | 0 | 70 | 0 | 0 | 2 | 81 | 81 | 1.04 |
| 2 HCHO+H4F→methylene-H4F | 103 | 0 | 100 | 100 | 0 | 0 | 0 | 0 | 2 | 71 | 0 | 3 | 0 | 0 | 0.00 |
| 3 Methenyl-H4F+NADPH↔Methylene-H4F+ NADP | 1.46 | 0.365 | 0.03 | 0.01 | 0 | 0.03 | 0.01 | 0 | 4 | 64 | 0 | 3 | 5 | 0 | 0.61 |
| 4 Methenyl-H4F↔Formyl-H4F | 1.55 | 2.86 | 0.08 | 0 | 0 | 0.08 | 0 | 0 | 4 | 0 | 0 | 6 | 0 | 0 | –0.61 |
| 5 Formate+ATP+H4F↔Formyl-H4F+ADP+Pi | 16.0 | 0.0352 | 22 | 0.021 | 0.08 | 0.08 | 0.021 | 2 | 7 | 8 | 71 | 6 | 65 | 76 | 0.61 |

| Reactions | $V_F$ | $V_B$ | $K_1$ | $K_2$ | $K_3$ | $K_1'$ | $K_2'$ | $K_3'$ | $i_1$ | $i_2$ | $i_3$ | $j_1'$ | $j_2'$ | $j_3'$ | s.s.f |
|---|---|---|---|---|---|---|---|---|---|---|---|---|---|---|---|
| 6 Formate+NAD→NADH+$CO_2$ | 0.804 | 0 | 1.6 | 0.07 | 0 | 0 | 0 | 0 | 7 | 1 | 0 | 63 | 73 | 0 | 0.43 |
| 7 HCHO+$H_4$MPT→Methylene-$H_4$MPT | 33.8 | 0 | 0.2 | 0.1 | 0 | 0 | 0 | 0 | 2 | 72 | 0 | 9 | 0 | 0 | 1.04 |
| 8 Methylene-$H_4$MPT+NAD↔Methenyl-$H_4$MPT+NADH | 0.876 | 0.0201 | 0.05 | 0.2 | 0 | 0.2 | 0.2 | 0 | 9 | 1 | 0 | 10 | 63 | 0 | 0.63 |
| 9 Methylene-$H_4$MPT+NADP↔Methenyl-$H_4$MPT+NADPH | 0.648 | 0.00745 | 0.1 | 0.02 | 0 | 0.2 | 0.02 | 0 | 9 | 5 | 0 | 10 | 64 | 0 | 0.42 |
| 10 Methenyl-$H_4$MPT↔formyl-$H_4$MPT | 0.372 | 18.1 | 0.03 | 0 | 0 | 0.2 | 0 | 0 | 10 | 0 | 0 | 11 | 0 | 0 | 1.04 |
| 11 Methanofuran+formyl-$H_4$MPT↔FMR+$H_4$MPT | 17.5 | 3.57 | 0.05 | 0.2 | 0 | 0.05 | 0.2 | 0 | 75 | 11 | 0 | 12 | 72 | 0 | 1.04 |
| 12 Formylmethanofuran→Formate | 6.26 | 0 | 0.05 | 0 | 0 | 0 | 0 | 0 | 12 | 0 | 0 | 7 | 0 | 0 | 1.04 |
| 13 Methylene-$H_4$F+glycine↔Serine+$H_4$F | 4.91 | 2.46 | 0.07 | 16.6 | 0 | 4.89 | 0.07 | 0 | 3 | 13 | 0 | 14 | 71 | 0 | 0.61 |
| 14 Serine+glyoxylate↔glycine+hydroxypyruvate | 4.95 | 2.48 | 4.89 | 12.6 | 0 | 16.6 | 1.4 | 0 | 14 | 15 | 0 | 13 | 16 | 0 | 0.62 |
| 15 Hydroxypyruvate+NADH↔glycerate+NAD+ | 4.95 | 2.48 | 1.4 | 0.53 | 0 | 2.22 | 1.69 | 0 | 16 | 63 | 0 | 17 | 1 | 0 | 0.62 |
| 16 Hdroxypyruvate+NADPH↔glycerate+NADP | 0 | 0 | 1.4 | 0.42 | 0 | 2.22 | 0.62 | 0 | 16 | 64 | 0 | 17 | 5 | 0 | 0 |
| 17 Glycerate+ATP→2-phosphoglycerate+ADP+Pi | 2.48 | 0 | 2.22 | 2.88 | 0 | 0 | 0 | 0 | 17 | 8 | 0 | 19 | 65 | 76 | 0.62 |
| 18 2-PG↔3-PG | 0.134 | 0.0668 | 2.19 | 0 | 0 | 2.36 | 0 | 0 | 19 | 0 | 0 | 18 | 0 | 0 | 0.03 |
| 19 2-PG↔phosphoenolpyruvate | 2.34 | 1.17 | 2.19 | 0 | 0 | 6 | 0 | 0 | 19 | 0 | 0 | 20 | 0 | 0 | 0.59 |
| 20 Phosphoenolpyruvate+$CO_2$↔Oxaloacetate+Pi | 1.46 | 0.731 | 6 | 0 | 0 | 7.34 | 0 | 0 | 20 | 73 | 0 | 21 | 76 | 0 | 0.37 |
| 21 Oxaloacetate+NADH→malate+NAD | 0.821 | 0 | 7.34 | 0.53 | 0 | 1.6 | 1.69 | 0 | 21 | 63 | 0 | 22 | 1 | 0 | 0.21 |
| 22 Malate+ATP+CoA→malyl-CoA+ADP+Pi | 21.1 | 0 | 1.6 | 2.88 | 0.1 | 3.43 | 0.6 | 3 | 22 | 8 | 26 | 23 | 65 | 76 | 0.62 |
| 23 Glyoxylate+Acetyl CoA↔Malyl-CoA | 1.49 | 2.98 | 12.61 | 0.38 | 0 | 3.43 | 0 | 0 | 15 | 24 | 0 | 23 | 0 | 0 | –0.62 |
| 24 Pyruvate+CoA+NAD→acetyl CoA+$CO_2$+NADH | 2.73 | 0 | 3.85 | 0.1 | 1.69 | 0 | 0 | 0 | 25 | 26 | 1 | 24 | 73 | 63 | 0.33 |
| 25 Acetyl CoA+OAA→citrate+CoA | 0.361 | 0 | 0.38 | 7.34 | 0 | 0 | 0 | 0 | 24 | 21 | 0 | 27 | 26 | 0 | 0.03 |
| 26 Cis-aconitate↔Citrate | 0.0556 | 0.111 | 1 | 0 | 0 | 4.74 | 0 | 0 | 28 | 0 | 0 | 27 | 0 | 0 | –0.03 |
| 27 Isocitrate↔cis-aconitate | 0.0556 | 0.111 | 1 | 0 | 0 | 1 | 0 | 0 | 29 | 0 | 0 | 28 | 0 | 0 | –0.03 |
| 28 Isocitrate+NADP↔a-KG+NADPH+$CO_2$ | 0.222 | 0.111 | 1 | 0.62 | 0 | 0.65 | 0.42 | 0 | 29 | 5 | 0 | 30 | 64 | 73 | 0.03 |
| 29 a-KG+NAD+CoA→succinyl CoA+$CO_2$+NADH | 0.124 | 0 | 0.65 | 1.69 | 0.1 | 0 | 0 | 0 | 30 | 1 | 26 | 31 | 73 | 63 | 0.01 |
| 30 Succinate+GTP+CoA↔succinyl CoA+Pi+GDP | 3.40 | 6.80 | 11.8 | 0.99 | 0.1 | 0.65 | 3 | 0.11 | 33 | 66 | 26 | 31 | 76 | 32 | –0.43 |
| 31 Fumarate+$FADH_2$↔Succinate+FAD | 1.71 | 3.43 | 0.51 | 0.11 | 0 | 11.8 | 0.99 | 0 | 34 | 67 | 0 | 33 | 35 | 0 | –0.43 |
| 32 Fumarate↔Malate | 1.70 | 0.85 | 0.51 | 0 | 0 | 1.6 | 0 | 0 | 34 | 0 | 0 | 22 | 0 | 0 | 0.43 |
| 33 Glc6P↔Fru6P | 0.00646 | 0.0129 | 4.2 | 0 | 0 | 6.36 | 0 | 0 | 36 | 0 | 0 | 37 | 0 | 0 | 0.00 |
| 34 Fru(1 6)$P_2$→Fru6P+Pi | 0.0304 | 0 | 2 | 0 | 0 | 6.36 | 0 | 0 | 38 | 0 | 0 | 37 | 76 | 0 | 0.01 |
| 35 Fru(1 6)$P_2$↔dihydroxyacteone phosphate+TP | 0.0390 | 0.0779 | 2 | 0 | 0 | 12.4 | 12.4 | 0 | 38 | 0 | 0 | 39 | 40 | 0 | –0.01 |
| 36 TP↔Dihydroxyacteone phosphate | 0.0432 | 0.0216 | 12.4 | 0 | 0 | 12.4 | 0 | 0 | 40 | 0 | 0 | 39 | 0 | 0 | 0.01 |
| 37 TP+Pi+NAD↔1 3)-DPG+NADH | 0.115 | 0.231 | 12.4 | 3 | 1.69 | 11.3 | 0.53 | 0 | 40 | 76 | 1 | 41 | 63 | 0 | –0.02 |
| 38 3-PG+ATP↔1 3)-DPG+ADP | 0.192 | 0.0959 | 2.36 | 2.88 | 0 | 11.3 | 0.6 | 0 | 18 | 8 | 0 | 41 | 65 | 0 | 0.02 |
| 39 OAA+GTP↔PEP+GDP+CO2 | 0.0823 | 0.0412 | 7.34 | 0.99 | 0 | 6 | 0.11 | 0 | 21 | 66 | 0 | 20 | 32 | 73 | 0.01 |
| 40 Pyruvate+$CO_2$+ATP↔OAA+ADP+Pi | 0.845 | 1.69 | 3.85 | 5 | 2.88 | 7.34 | 0.6 | 3 | 25 | 73 | 8 | 21 | 65 | 76 | -0.11 |
| 41 PEP+ADP→pyruvate+ATP | 1.79 | 0 | 6 | 0.6 | 0 | 0 | 0 | 0 | 20 | 65 | 0 | 25 | 8 | 0 | 0.23 |
| 42 Malate+NAD→pyruvate+$CO_2$+NADH | 0.804 | 0 | 1.6 | 1.69 | 0 | 0 | 0 | 0 | 22 | 1 | 0 | 25 | 73 | 63 | 0.01 |

| Reactions | $V_F$ | $V_B$ | $K_1$ | $K_2$ | $K_3$ | $K_1'$ | $K_2'$ | $K_3'$ | $i_1$ | $i_2$ | $i_3$ | $j_1'$ | $j_2'$ | $j_3'$ | s.s.f |
|---|---|---|---|---|---|---|---|---|---|---|---|---|---|---|---|
| 43 Glc6P+NADP→6-PG+NADPH | 0.0397 | 0 | 4.2 | 0.62 | 0 | 0 | 0 | 0 | 36 | 5 | 0 | 42 | 64 | 0 | 0.00 |
| 44 6-PG+NADP→ribose 5-phosphate+$CO_2$+NADPH | 0.0129 | 0 | 2 | 0.62 | 0 | 0 | 0 | 0 | 42 | 5 | 0 | 43 | 73 | 64 | 0.00 |
| 45 Ribose 5-phosphate↔ribulose 5-phosphate | 0.00236 | 0.00472 | 4.51 | 0 | 0 | 4.51 | 0 | 0 | 43 | 0 | 0 | 44 | 0 | 0 | 0.00 |
| 46 Ribulose 5-phosphate↔xylulose 5-phosphate | 0.00236 | 0.00472 | 4.51 | 0 | 0 | 1.8 | 0 | 0 | 44 | 0 | 0 | 45 | 0 | 0 | 0.00 |
| 47 Xylulose 5-phosphate+ribose 5-phosphate↔$S_7$P+TP | 0.00327 | 0.00163 | 1.8 | 4.51 | 0 | 1.85 | 12.4 | 0 | 45 | 43 | 0 | 46 | 40 | 0 | 0.00 |
| 48 S7P+TP↔Fru-6P+Ery-4P | 0.00329 | 0.00164 | 1.85 | 12.4 | 0 | 6.36 | 3.59 | 0 | 46 | 40 | 0 | 37 | 47 | 0 | 0.00 |

| # | Reaction | $V_F$ | $V_B$ | $K_1$ | $K_2$ | $K_3$ | $K'_1$ | $K'_2$ | $K'_3$ | $i_1$ | $i_2$ | $i_3$ | $j_1$ | $j_2$ | $j_3$ | s.s.f |
|---|---|---|---|---|---|---|---|---|---|---|---|---|---|---|---|---|
| 49 | Xylulose 5-phosphate+Ery-4P↔Fru-6P+TP | 0.00636 | 0.0127 | 1.8 | 3.59 | 0 | 6.36 | 12.4 | 0 | 45 | 47 | 0 | 37 | 40 | 0 | 0.00 |
| 50 | 2Acetyl CoA↔Acetoac CoA+CoA | 3.64 | 1.82 | 0.38 | 0.38 | 0 | 0.18 | 0.1 | 0 | 24 | 24 | 0 | 48 | 26 | 0 | 0.45 |
| 51 | Acetoac CoA+NADPH↔3HB-CoA+NADP | 2.88 | 1.44 | 0.18 | 0.42 | 0 | 0.03 | 0.62 | 0 | 48 | 64 | 0 | 49 | 5 | 0 | 0.42 |
| 52 | 3HB-CoA↔PHB+CoA | 0.400 | 0.200 | 0.03 | 0 | 0 | 0.12 | 0.1 | 0 | 49 | 0 | 0 | 50 | 26 | 0 | 0.06 |
| 53 | PHB→3HB | 0.0508 | 0 | 0.12 | 0 | 0 | 1.16 | 0 | 0 | 50 | 0 | 0 | 51 | 0 | 0 | 0.03 |
| 54 | 3HB+NAD↔Acetoac+NADH | 0.204 | 0.102 | 1.16 | 1.69 | 0 | 0.85 | 0.53 | 0 | 51 | 1 | 0 | 52 | 63 | 0 | 0.03 |
| 55 | Acetoac+succinyl CoA↔Acetoac CoA+succinate | 0.0322 | 0.0644 | 0.85 | 0.65 | 0 | 0.18 | 11.8 | 0 | 52 | 31 | 0 | 48 | 33 | 0 | -0.01 |
| 56 | Crot-Coa↔3HB-Coa | 2.67 | 5.34 | 0.01 | 0 | 0 | 0.03 | 0 | 0 | 53 | 0 | 0 | 49 | 0 | 0 | -0.36 |
| 57 | L3HB-Coa↔Crot-Coa | 0.404 | 0.202 | 0.24 | 0 | 0 | 0.01 | 0 | 0 | 54 | 0 | 0 | 53 | 0 | 0 | 0.05 |
| 58 | L3HB-Coa+NAD↔Acetoac CoA+NADH | 0.206 | 0.412 | 0.24 | 1.69 | 0 | 0.18 | 0.53 | 0 | 54 | 1 | 0 | 48 | 63 | 0 | -0.05 |
| 59 | Crot-Coa+NADPH↔but-COA+NADP | 2.04 | 1.02 | 0.01 | 0.42 | 0 | 0.22 | 0.62 | 0 | 53 | 64 | 0 | 55 | 5 | 0 | 0.41 |
| 60 | But-CoA+FAD+2NAD→prop-CoA+FADH+2NADH+ $CO_2$ | 6.62 | 0 | 0.22 | 0.99 | 1.69 | 0 | 0 | 0 | 55 | 35 | 1(1) | 56 | 67 | 63 | 0.41 |
| 61 | Prop-CoA+$CO_2$+ATP↔Mema-CoA+ADP+Pi | 5.94 | 2.97 | 0.09 | 5 | 2.88 | 0.27 | 0.6 | 0 | 56 | 73 | 8 | 57 | 65 | 76 | 0.41 |
| 62 | Mema-CoA↔Succ-CoA | 1.65 | 0.82 | 0.27 | 0 | 0 | 0.65 | 0 | 0 | 57 | 0 | 0 | 31 | 0 | 0 | 0.41 |
| 63 | 3-PG+NAD↔PHP+NADH | 0.0756 | 0.0378 | 2.36 | 1.69 | 0 | 0.6 | 0.53 | 0 | 18 | 1 | 0 | 58 | 63 | 0 | 0.01 |
| 64 | PHP+Glutamate↔a-KG+3Pser | 0.0715 | 0.0357 | 0.6 | 12 | 0 | 0.65 | 0.09 | 0 | 58 | 60 | 0 | 30 | 59 | 0 | 0.01 |
| 65 | 3Pser→serine+Pi | 0.0192 | 0 | 0.09 | 0 | 0 | 0 | 0 | 0 | 59 | 0 | 0 | 14 | 76 | 0 | 0.01 |
| 66 | Glutamate+NADP←NH4+a-KG+NADPH | 0 | 0.316 | 12 | 0.62 | 0 | 1 | 0.65 | 0.42 | 60 | 5 | 0 | 74 | 30 | 64 | -0.02 |
| 67 | NADH+Q+H->NAD+$QH_2$ | 47.9 | 50 | 0.53 | 0.98 | 6.3 | 1.69 | 1.02 | 0 | 63 | 61 | 77 | 1 | 68 | 0 | -0.30 |
| 68 | $QH_2$+2Cyt-cox↔Q+2Cyt-cred+2H | 513 | 500 | 1.02 | 0 | 0 | 0.98 | 6.3 | 6.3 | 68 | 0 | 0 | 61 | 77 | 77 | 1.72 |
| 69 | $FADH_2$+Q↔FAD+$QH_2$ | 23.1 | 20 | 0.11 | 0.98 | 0 | 0.99 | 1.02 | 0 | 67 | 61 | 0 | 35 | 68 | 0 | 0.84 |
| 70 | ATP↔ADP+2H+Pi | 191 | 200 | 2.88 | 0 | 0 | 0.6 | 6.3 | 6.3 | 8 | 0 | 0 | 65 | 77 | 77 | -1.80 |
| 71 | ATP+GDP↔ADP+GTP | 23.1 | 25 | 2.88 | 0.11 | 0 | 0.6 | 0.99 | 0 | 8 | 32 | 0 | 65 | 66 | 0 | -0.42 |
| 72 | NADPH+Q+H↔NAPD+$QH_2$ | 50.5 | 50 | 0.42 | 0.98 | 6.3 | 0.62 | 1.02 | 0 | 64 | 61 | 77 | 5 | 68 | 0 | 0.14 |
| 73 | NADH+NADP↔NADPH+NAD | 105 | 100 | 0.53 | 0.62 | 0 | 0.42 | 1.69 | 0 | 63 | 5 | 0 | 64 | 1 | 0 | 1.26 |
| 74 | 2e+Q→$QH_2$ | 8.33 | 0 | 0.2 | 0.2 | 0.98 | 0 | 0 | 0 | 81 | 81 | 61 | 68 | 0 | 0 | 1.04 |
| 75 | 2Glyoxylate↔hydropyruvate | 0 | 0 | 12.6 | 12.6 | 0 | 1.4 | 0 | 0 | 15 | 15 | 0 | 16 | 0 | 0 | 0 |
| 76 | Serine+NH4↔Pyruvate | 0.00541 | 0.00271 | 4.89 | 0 | 0 | 3.85 | 0 | 0 | 14 | 0 | 0 | 25 | 0 | 0 | 0.00 |
| 77 | Acetoac+CoA+ATP↔Acetoac-CoA+ADP+Pi | 0.544 | 0.272 | 0.85 | 0.1 | 2.88 | 0.18 | 0.6 | 3 | 52 | 26 | 8 | 48 | 65 | 76 | 0.03 |
| 78 | Glycine+NAD+$H_4F$↔Methylene-$H_4F$+NADH+$NH_3$+$CO_2$ | 0.000431 | 0.000216 | 16.6 | 1.69 | 0 | 0.07 | 0.53 | 0 | 13 | 1 | 0 | 3 | 63 | 0 | 0.00 |
| 79 | Pyruvate-Out↔Pyruvate | 5.65 | 0.01 | 10 | 0 | 0 | 10 | 0 | 0 | 78 | 0 | 0 | 25 | 0 | 0 | 0.00 |
| 80 | Succinate-Out↔Succinate | 7.00 | 0.25 | 0.1 | 0 | 0 | 10 | 0 | 0 | 79 | 0 | 0 | 33 | 0 | 0 | 0.00 |

Columns $V_F$ and $V_B$ are the forward and backward velocity (in millimolar/second) of the reactions respectively. Columns $K_{i=1,2,3}$ are the (Michalis-Menten-like) $K_m$ (in millimolar) of the reactants. Columns $K'_{j=1,2,3}$ are the (Michaelis-Menten-like) $K_m$ (in millimolar) of the products. Columns $i_1, i_2, i_3$ contains the metabolite index for the reactants. Columns $j_1, j_2, j_3$ contains the metabolite index for the products. Column s.s.f stands for steady state flux (in millimolar/second)

Next we show a result from perturbing the steady state. We set the initial concentrations at the steady state value except for one. In this case, the formaldehyde concentration is set to twice it's steady state value. We follow the predicted dynamics of the system and see how the concentrations relax back to their steady state value. In Fig. 2., we plot the predicted concentrations divided by their steady state values as a function of time. Each line represents a metabolite's concentration in the plot and in this plot we have only labeled a few that show a large relative variation. We find the system is predicted to be robust, as it relaxes back to the same steady state eventually. Different metabolites also relax back to the steady state at different times. Such behaviors are consistent with in vivo experimental observations.

Since formaldehyde is highly toxic to most organisms, it is of great interest to understand how the bacterium deals with such a toxic intermediate[55]. Our simulation predicts that the formaldehyde concentration is able to reach steady state concentration in the shortest time as compared to other metabolites (e.g. succinate not shown). This suggests that a mechanism is already in place in our kinetic model to deal with sudden large fluctuations in formaldehyde concentrations. Furthermore, perturbation in formaldehyde concentration is not predicted to impact other pathways as much. Fig. 3 shows the corresponding predicted variations in the fluxes. Again, such dynamical behaviors are consistent with biological observations.

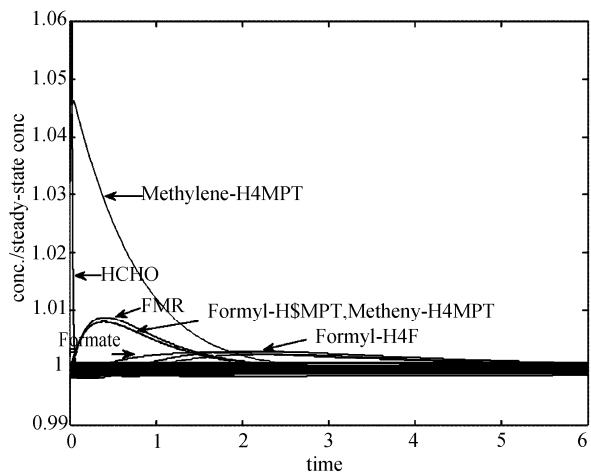

**Fig. 2 Response in concentration to a perturbation in the formaldehyde concentration at steady state values**

The concentration of formaldehyde was doubled at $t=0$ while the other concentration are at steady state values

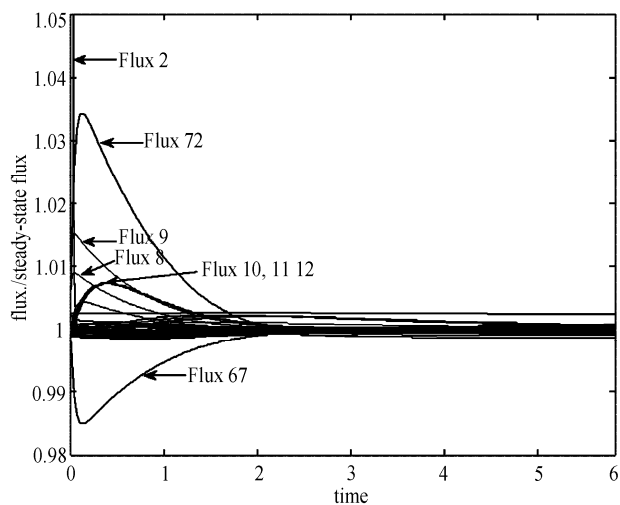

**Fig. 3 Response in flux to a perturbation in the formaldehyde concentration at steady state values**

The concentration of formaldehyde was doubled at $t=0$ while the other concentration are at steady state values. The numbering of fluxes is in accordance with numbering of reactions as in Table 3

## 3 Conclusions

We have been developing a systematic methodology to perform the real time kinetic modeling of large metabolic networks based on incompletely kinetic parameter information. A successful major step is reported here: The validation of the modeling has been carried out using the central metabolism of *M. extorquens* AM1 under methanol growth.

The critical issue we encountered in constructing such computational model is the large number of unknown parameters in the enzymatic rate equations. We use generic rate equations which can be written with a minimum set of parameters. Some kinetic parameters can be directly inferred from experimental values, and directly from those of concentrations and fluxes. The rest can be computed according to the procedure described in the present paper.

We also demonstrate that the model is robust with respect to fluctuations in formaldehyde in Fig. 2 where formaldehyde concentration is doubled from its steady state value. In our case, equilibrium is reached by the mathematical model via an autonomous adjusting substrate intake and its biomass production according to the need for metabolites and energy. A full exploration of other situations, such as different substrates, gene knockouts, will be explored in following-up publications.

**Acknowledgements:** We acknowledge Ludmila Chistoserdova, Greg Crowther, Marina Kalyuzhnaya, George Kosály, Yoko Okubo, Elizabeth Skovran, Stephen J. Van Dien for valuable discussions through this project.